# Bright, Coherent, Ultrafast Soft X-Ray Harmonics Spanning the Water Window from a Tabletop Light Source

M.C. Chen, P. Arpin, T. Popmintchev, M. Gerrity, B. Zhang, M. Seaberg, M.M. Murnane and H.C. Kapteyn

*JILA, University of Colorado at Boulder, Boulder, CO 80309-0440*

*Ph. (303) 210-0396; FAX: (303) 492-5235; E-mail: murnane@jila.colorado.edu*

**Abstract**

We demonstrate fully phase matched high-order harmonic generation with emission spanning the water window spectral region important for bio- and nano-imaging and a breadth of materials and molecular dynamics studies. We also generate the broadest bright coherent bandwidth (≈ 300eV) to date obtained from any light source, small or large. The harmonic photon flux at 0.5 keV is $10^3$ higher than demonstrated previously, making it possible for the first time to demonstrate spatial coherence in the water window. The continuum emission is consistent with a single attosecond burst, that extends bright attosecond pulses into the soft x-ray region.

Nonlinear optics has revolutionized laser science by making it possible to efficiently convert laser light from one wavelength to another. Using the extreme nonlinear-optical process of high harmonic generation (HHG), light from an ultrafast laser can be coherently upshifted, resulting in a useful, tabletop, coherent short wavelength source. However, despite the fact that harmonics in the water window were first observed in 1997 [1, 2] most HHG applications to-date have used EUV and not soft x-ray light, due to the limited flux there. The unique properties of HHG have already proven useful for studying ultrafast molecular, plasma and materials dynamics [3-5], for characterizing heat flow on nanoscale dimensions [6], for following element-specific dynamics in magnetic materials [7], and for high resolution coherent imaging [8-10]. Moreover, the attosecond recollision physics that underlies HHG produces femtosecond-to-attosecond pulses that are ideal for capturing the motion of electrons in atoms, molecules and materials on their fundamental time (~<fs) and length (~<nm) scales.

In HHG, short wavelength light is emitted as a result of the highly nonlinear motion of an electron in the process of ionizing from an atom in a strong laser field. The maximum HHG photon energy emitted by a *single atom* is given by a simple cutoff rule: $h\nu_{max} = I_p + 3.2 U_p$, where $I_p$ is the ionization potential of the atom and $U_p \approx I_L \lambda_L^2$ is the quiver energy of an electron in a laser field of intensity $I_L$ and wavelength $\lambda_L$, where I is the laser intensity at which the atom is field ionized [11, 12]. This very favorable scaling for hν$_{max}$ means that weak harmonics can be generated up to the water window soft x-ray spectral region (between the C and O K-shell absorption edges at 284 - 540 eV) [1, 2] and even into the keV hard x-ray region of the spectrum but with very low flux levels [13]. For efficient frequency upconversion however, the generated harmonics must propagate at the same phase velocity as the driving laser, allowing for phase-matched, coherent growth of the signal from many atoms in an extended nonlinear medium. Full phase matching of the HHG process requires that dispersion due to the neutral gas, plasma, quantum, and any geometrical contribution be balanced. In a hollow waveguide geometry [14-16], phase matching ($\Delta k = k_{q\omega} - qk_{\omega} = 0$) can be achieved at some optimum pressure $p$ given by –

$$\Delta k \approx q \underbrace{\frac{u_{11}^2 \lambda_L}{4\pi a^2}}_{geometric} - \underbrace{qp(1-\eta)\frac{2\pi}{\lambda_L}(\Delta\delta + n_2)}_{atoms} + \underbrace{qp\eta N_a r_e \lambda_L}_{free\ electrons}. \quad (1)$$

Here $q$ denotes the harmonic order, $u_{11}$ is the mode factor, $a$ is the radius of the waveguide, $\eta$ is the ionization level, $r_e$ is the classical electron radius, $N_a$ is the number density of atoms per atm, $\Delta\delta$ is the difference between the indices of refraction of the gas at the fundamental and x-ray wavelengths, $\omega$ and $\lambda_L$ are the fundamental laser frequency and wavelength respectively, $k$ is the phase velocity, and $n_2 = \tilde{n}_2 I_L$ is the nonlinear index of refraction at $\lambda_L$.

However, from Eqn. 1, phase-matching of the HHG process is possible only for ionization levels below a critical level of ionization given by $\eta_c(\lambda_L) = [1 + \lambda_L^2 r_e N_{atm} / (2\pi\Delta\delta(\lambda_L))]^{-1}$.

Because higher harmonics are emitted at higher levels of ionization in the medium [17], *this results in a phase matching cutoff [18, 19] that is always lower than the single atom cutoff proportional to* $\approx I \lambda_L^2$. Physically, when the ionization in the gas exceeds $\eta_{cr}(\lambda_L)$, the phase velocity of the driving laser exceeds that of the generated harmonics, and phase matching is not possible even when using very short driving laser pulses. For Ti:sapphire driving lasers at a wavelength of 0.8 μm, this limits bright harmonic emission to photon energies < 150 eV. Overcoming this phase matching limit has thus become a grand challenge in nonlinear optics, that has motivated a variety of schemes such as quasi phase matching (QPM) [20-23], non-adiabatic and short pulse phase matching [13], and neutral atom phase matching [24]. However, to date none of these schemes has succeeded in generating bright coherent harmonics over the extended (absorption-limited) propagation length necessary to maximize conversion efficiency.

Very recently we showed that by using longer wavelength, mid-infrared (mid-IR) driving lasers, full phase matching of the HHG process can be extended in theory to >10 keV [18, 19]. The single atom cutoff rule, $h\nu_{max} \approx I \lambda_L^2$, clearly shows that using a longer wavelength driving laser will result in the generation of shorter wavelength harmonics. A number of recent studies have borne this out [25, 26]. However, these studies also showed a dramatically decreasing photon yield[27-29] with longer driving laser wavelengths, scaling as $\approx \lambda^{-6.5}$. Fortunately, from the cutoff rule, to generate a given energy harmonic, the required laser intensity is significantly lower for longer driving laser wavelengths. This reduces ionization in the medium, and extends full phase matching to much higher photon energies. In past work we demonstrated full phase matching of HHG in He at photon energies up to 330 eV, that just reached the low edge of the water window region important for applications [18]. Other recent work demonstrated phase matching over ≈ mm regions of a supersonically expanding neutral gas at photon energies around 270 eV in Ne (with weak emission extending to 400 eV in He) [24].

In this paper we extend bright, spatially coherent, high harmonic emission throughout the water window, to photon energies > 0.5 keV for the first time. The phase matched high harmonic flux is >$10^3$x brighter than previously demonstrated, and sufficient for a range of applications. This high flux also makes it possible for the first time to demonstrate spatial coherence of a compact source in the water window. Because of broad phase matching conditions, we also generate the broadest bright coherent bandwidths to date from any light source, spanning a 300eV soft x-ray supercontinuum. Finally, through careful comparison of theory and experiment, we validate that the same phase matching mechanism that operates for lower energy EUV harmonics driven by 0.8 µm light also works in the soft x-ray region around 0.5 keV when driven by mid-IR light i.e. by matching the phase velocity of the laser to the harmonics, the maximum high harmonic flux, limited only by gas absorption, can be achieved. The demonstration of bright, fully phase matched, coherent, high harmonic beams in the water window is significant for applications in bio-, molecular, nano- and materials imaging, where the presence of many elemental absorption edges enable high contrast, element-specific, studies with unprecedented spatial and temporal resolution.

In our experiment, 40 fs pulses at a wavelength of 2 μm (6 cycles FWHM) are generated in a three-stage optical parametric amplifier (OPA) seeded by a white-light continuum. High harmonics are then generated in He, Ne and Ar by focusing 2.4 mJ of the 2 μm idler beam into a 200 μm radius, 1 cm long, gas-filled hollow waveguide [18, 19]. The HHG spectrum is then detected using a flat-field imaging x-ray spectrometer and an x-ray CCD camera. Various metal filters are used to eliminate the fundamental laser light and to calibrate the spectrometer, depending on the photon energy range under investigation (e.g. B (K-edge 188 eV), Sc (L edge 400 eV), Ti (L edge 454 eV), and Cr (L edge 575 eV)).

Figure 1 plots the observed and predicted (from Eqn. 1) phase matching cutoffs for HHG in He, Ne and Ar using mid-IR driving pulses, for this work using 2 μm driving lasers, and for previous work using shorter wavelengths [18, 19, 24]. In He driven by 2 μm light, the phase matching cutoff (where the signal drops 50% from the peak) extends up to 520eV, while the harmonic emission peaks around 450 eV, extending past the O K-edge at 540 eV. The phase matched HHG signal around 450 eV also grows quadratically with pressure and begins to saturate at pressures of 8000 torr, as expected for phase matched HHG emission in the presence of absorption. In Ne, the phase matching cutoff extends to 395 eV and peaks at 370 eV at a pressure of 2600 torr, while in Ar, the cutoff extends to 165 eV and peaks at 145 eV, at a pressure of 800 torr. (The critical ionization levels are 0.0788%, 0.152%, and 0.628% in He, Ne and Ar driven by 2 μm light.) The observed phase matching pressures are also in excellent agreement with the predicted values from Eqn. 1, as shown in Fig. 2. Finally, from Eqn. 1 and Fig. 1, we can derive a simple scaling of $\lambda_L^{1.4-1.7}$ for the highest energy harmonic that can be phase matched in the soft x-ray region. This is nearly as favorable as the single atom cutoff rule.

The data of Figs. 1 and 2 clearly show that the observed phase matching cutoffs and pressures are in excellent agreement with theory, for mid-IR phase matching based on Eqn. 1. This finding is significant because other schemes have been proposed for generating bright harmonics at high photon energies - such as non-adiabatic phase matching [30, 31], reduced phase mismatch using very intense few-cycle driving laser pulses [2], or neutral gas phase matching [24]. However, to date these schemes have not resulted in bright harmonics in the water window. Moreover, the first two of these schemes are not general phase matching schemes, while the third scheme is fundamentally the same physics as discussed here. Our combined theoretical and experimental results thus validate that the same simple approach of dispersion control for phase matching, that generates bright harmonics in the EUV using 0.8 μm driving lasers, also works in the soft X-ray region for mid-IR driving lasers - provided that the gas pressure is increased by orders of magnitude to mitigate the low single atom emission. Finally, our data also show that pulses shorter than 10 optical cycles are not required for generating bright, ultrafast, harmonics in the soft x-ray region.

An important question to address is the brightness of the phase matched harmonic emission is in the water window region of the spectrum. Figure 3(a) plots the fully phase matched HHG flux from He and Ne, which is nearly constant on a linear scale from 250 eV to 520 eV. The HHG flux estimates are based on two approaches. First, we imaged the soft x-ray beam on the CCD camera, and derived the flux based on the counts, using the known quantum efficiency and gain of the camera and the filter transmissions. Second, we compared the water window emission with the phase matched HHG flux at 0.8 μm and 1.3 μm, where the HHG flux using 0.8μm kHz lasers was measured using a NIST calibrated vacuum photodiode. Based on these measurements, an approximate brightness of $\approx 10^6$ photons/sec or $\approx 10^5$ photons/shot is observed in a 1% bandwidth around 450 eV in He (at a repetition rate of 10 Hz). Moreover, a total of $6 \times 10^7$ photons/s is observed over the entire water window region (284 – 520 eV). These flux levels are $> 10^3$ higher than achieved to date at 0.5 keV. Further increases in flux by many orders of magnitude can be expected by using driving lasers with better beam quality and higher repetition rates, higher He pressures and longer waveguides with better differential pumping.

To demonstrate that the HHG beam is spatially coherent [32], we implemented the first coherence measurement in the water window using any compact light source. We placed a double slit in the beam ~30 cm after the waveguide, followed by an x-ray CCD camera placed ~1 m after the double slit. The width of each slit was 10 μm with a center-to-center separation of 20 μm. The entire 60eV FWHM bandwidth harmonic emission from Ne (shown in Fig. 4 (a)) was used to illuminate the double slit, with a beam waist of 1.7mm. Figure 4(a) shows the measured and predicted interference patterns, while Fig. 4(b) shows a lineout of the measured diffraction intensity (red) and simulated lineout (black) for a central wavelength of 3.8 nm (photon energy of 330 eV, with a bandwidth $\Delta E$ = 60 eV). The broad spectral bandwidth limits the fringe visibility off-center, as expected. The strong modulations of the measured diffraction pattern, combined with the excellent agreement between the simulations and data, demonstrate that the beam center is spatially coherent. Notice that in the image, only the 1[st] and 3[rd] fringes are visible. The 2[nd] fringe is not visible due to overlapping interferences between the single slit and double slit diffraction patterns. The low divergence of the 3nm light prevented us from characterizing the entire beam, because the fringes become smaller than the CCD pixels as the slits are separated, and the diffraction patterns from the pinholes do not overlap unless the pinhole size is reduced.

The phase matched harmonic emission in He, at > 300 eV, is the broadest bright coherent bandwidth generated to date from any light source (see Fig. 3(a)) and extends bright attosecond pulses into the soft x-ray region for the first time. Past theoretical work has shown that because phase matching is confined to only a few half-cycles of the laser - even when using relatively long 6-cycle driving laser pulses (15 fs at 800 nm or 40 fs at 2 µm) - the harmonics emerge as a single attosecond burst. This prediction was recently confirmed using 15 fs, 800 nm, driving lasers, where chirped pulses as short as 200 as were measured [33]. Here, the 2 µm idler beam from the OPA is intrinsically phase stabilized, which was verified experimentally (see Fig. 3(b) inset). Thus, when combined with a reduction of the attochirp ($1/\lambda_L$) with longer driving lasers, together with chirp compensation techniques that have been demonstrated previously [34], there is an excellent potential for generating pulses as short as $11 \pm 1$ as, which is the transform-limited bandwidth of the HHG emission from He, assuming a flat phase.

In summary, we demonstrated bright, coherent high harmonic emission throughout the water window region of the spectrum for the first time using any compact light source. Because of broad phase matching conditions, we observe the broadest bright coherent bandwidths from any light source to date, spanning 300 eV. We also validated a mechanism for phase matching the high harmonic upconversion process at high photon energies: the same dispersion balance mechanism that applies in the EUV region using 0.8 µm lasers also applies in the soft X-ray region using mid-IR driving laser wavelengths, provided the gas pressure is increased by orders of magnitude. Finally, we implemented spatial coherence measurements in the soft x-ray region for the first time using any compact light source. These advances will enable high-resolution tabletop microscopies of materials, nano- and biological samples with unprecedented combined spatial and temporal resolution.

The authors gratefully acknowledge funding from an NSSEFF Fellowship and from the NSF Engineering Research Center in EUV Science and Technology.

**Figure Captions**

**Figure 1:** (a) Solid (dashed) color lines show the predicted full phase-matching-cutoff energies as a function of the driving laser wavelength, for a pulse duration of three (eight) optical cycles. Solid circles show the observed full phase matching cutoff at 1.3 $\mu$m and 1.55 $\mu$m from [18] and [24], with the results of this paper plotted for 2 $\mu$m. Vertical stripes show the observed phase matching bandwidths. (b) Experimental pressure-tuned HHG spectra as a function of pressure using 2 $\mu$m driving pulses. The predicted phase matching cutoffs are shown as dashed lines.

**Figure 2:** Comparison between the measured (circles) and predicted (dashed lines based on Eqn. 1) optimal phase matching pressures at the phase matched cutoff photon energy in He, Ne and Ar, driven by 0.8$\mu$m, 1.3$\mu$m and 2$\mu$m lasers. The calculations assume ionization levels between 96% and 98% of critical ionization.

**Figure 3:** (a) Flux comparison between the ultra-broad bandwidth, phase-matched, HHG flux from He and Ne driven by 2 μm lasers, together with the transmission curves of the filters and gas. The dip in the spectrum around 284 eV is due to carbon contamination. (b) Fourier Transform of the HHG spectra from He at a pressure of 9000 torr, assuming a flat phase, showing the potential to support an 11±1 attosecond pulse, provided that the chirp is compensated. **(inset)** An f-2f interference measurement (each data point is averaged over 10 laser shots) shows long term

stability of the 2 μm light, with an rms jitter of the carrier wave with respect to the pulse envelope of 210 mrad over 40 mins.

**Figure 4: (a)** Measured and calculated double-slit interference pattern using HHG from Ne at 330 eV. **(b)** Comparison between the measured (red) and simulated diffraction pattern lineout (black) assuming a central wavelength of 3.8 nm (330 eV). The broad bandwidth ($\Delta E = 60$ eV) limits the number of fringes. One fringe is not visible due to interference between the diffraction from a single and double slit.

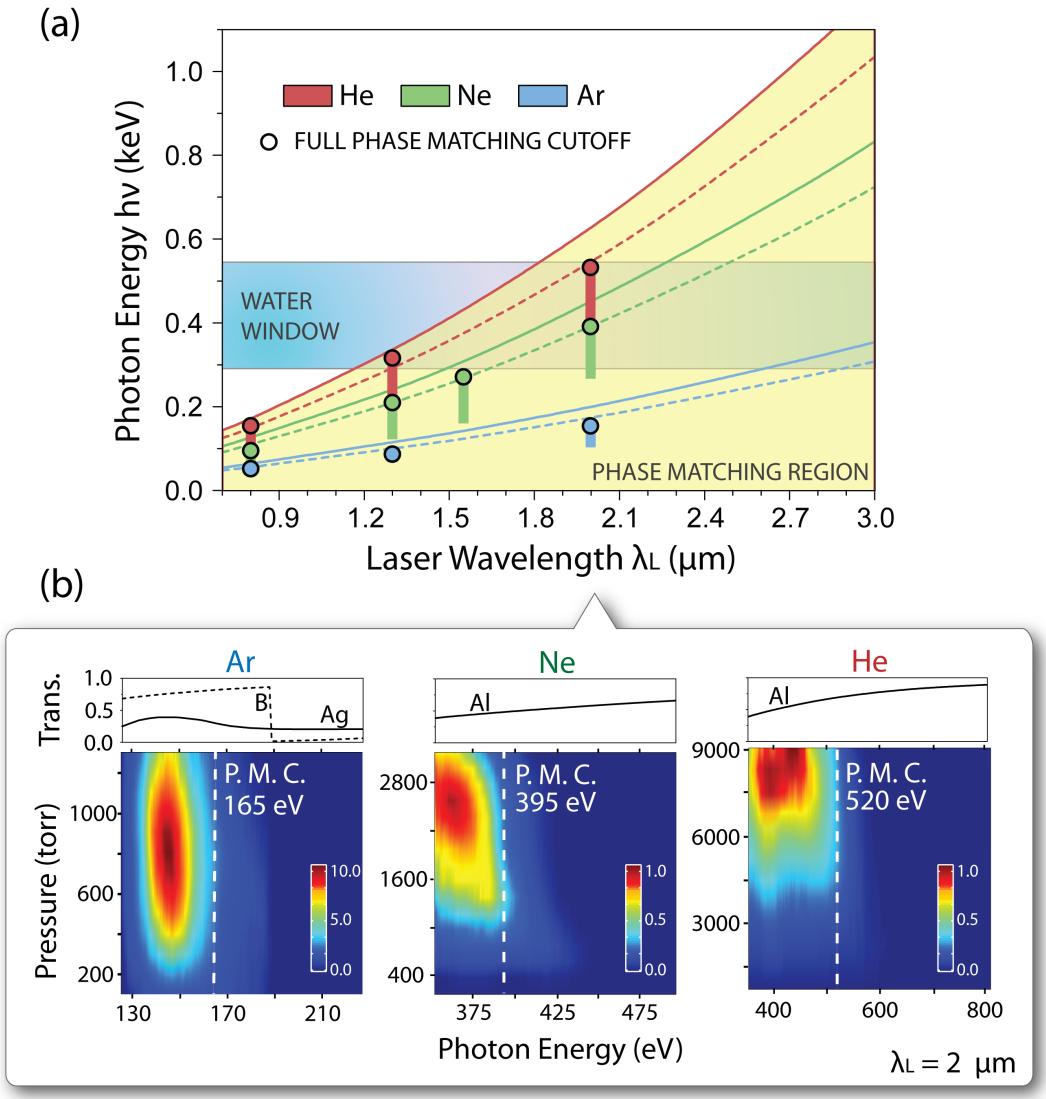

Figure 1

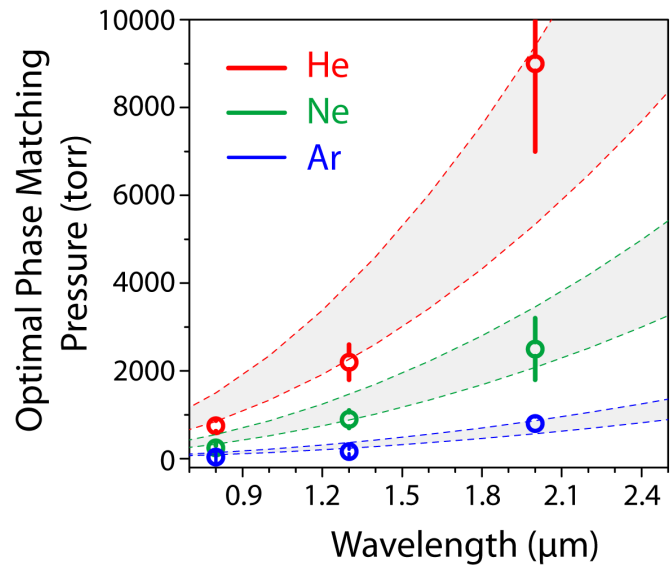

**Figure 2**

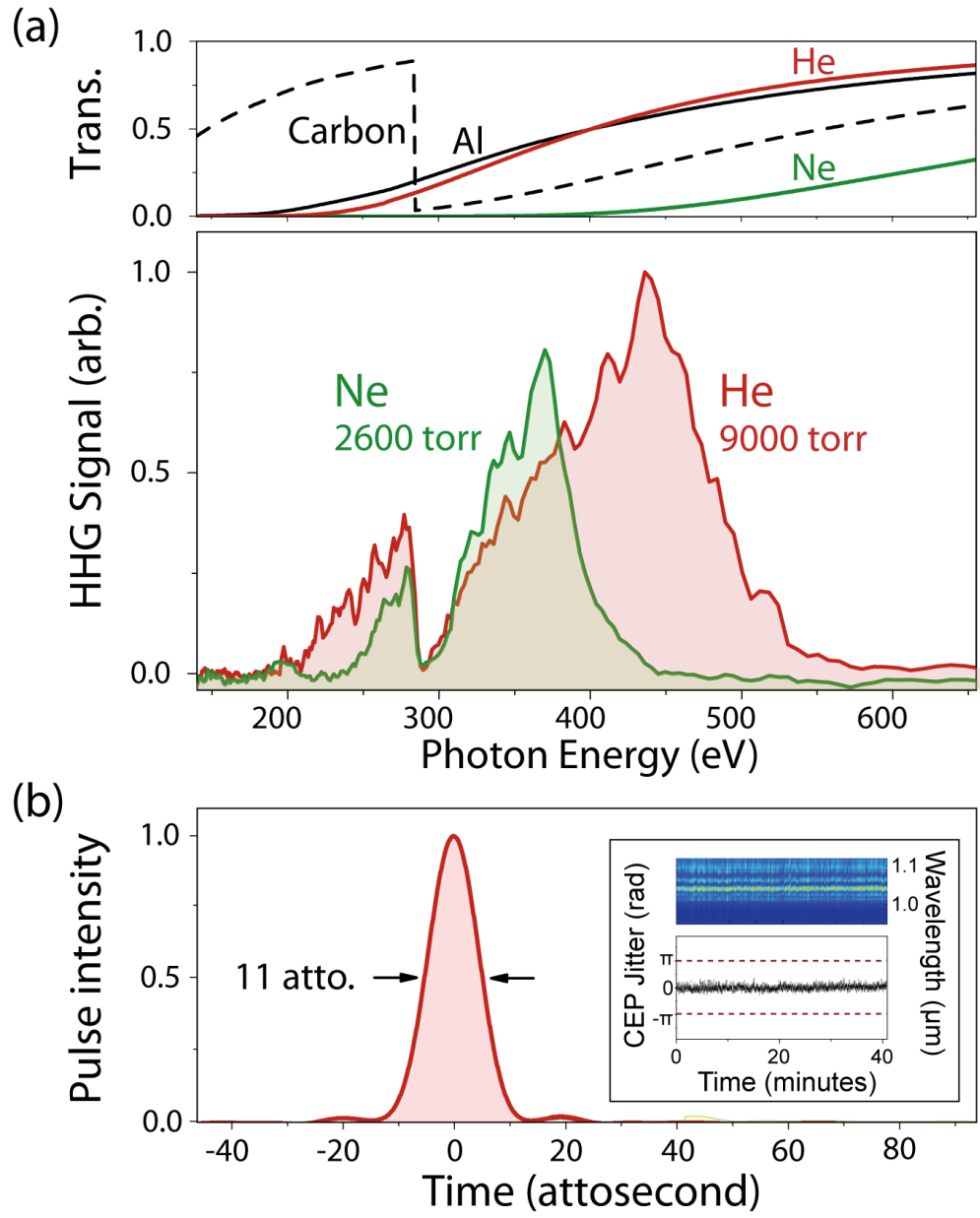

**Figure 3**

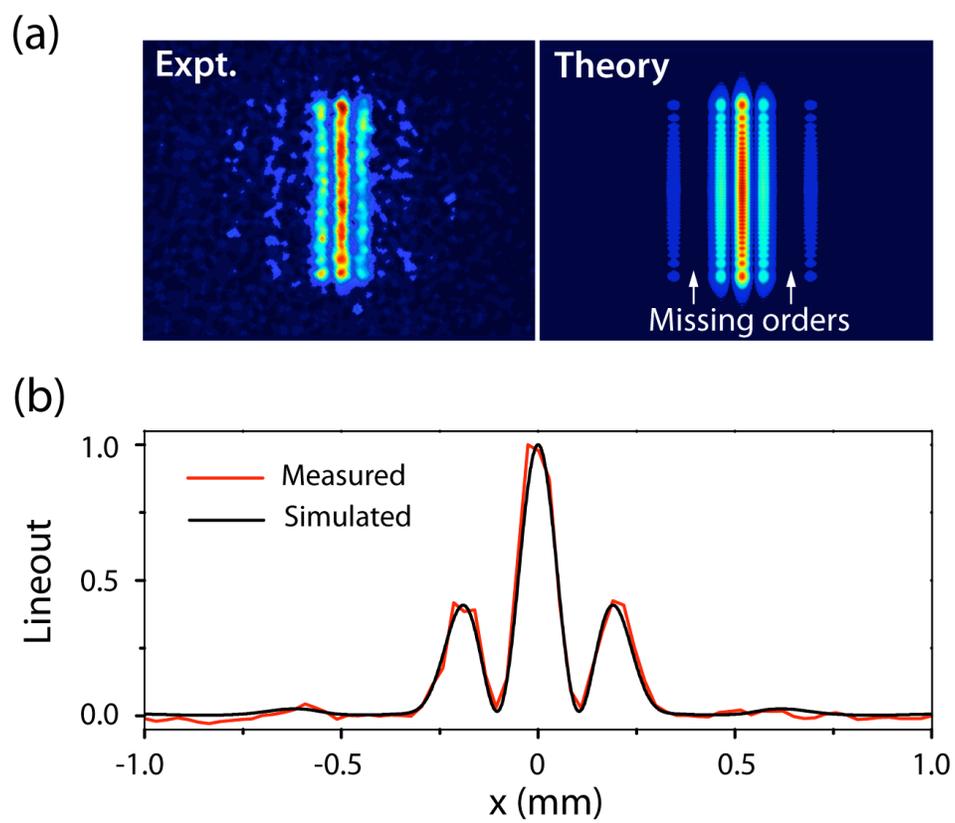

**Figure 4**